\documentclass{pasj00}
\draft

\begin{document}
\SetRunningHead{K. Asano and S. Kobayashi}
{Dispersion of the Break Energy in the GRB Internal Shock Model}
\Received{2003/01/8}
\Accepted{2003/03/19}

\title{Dispersion of the Break Energy in the GRB Internal Shock Model}
\author{Katsuaki \textsc{Asano}}
\affil{Department of Earth and Space Science,
Osaka University, Toyonaka, Osaka 560-0043}
\email{asano@vega.ess.sci.osaka-u.ac.jp}
\and
\author{Shiho \textsc{Kobayashi}}
\affil{Center for Gravitational Wave Physics, Pennsylvania State University, \\
University Park, PA 16802, U.S.A.}
\affil{Department of Astronomy and Astrophysics, Pennsylvania State University, \\
University Park, PA 16802, U.S.A.}
\email{shiho@gravity.psu.edu}

\KeyWords{gamma rays: bursts---radiation mechanisms:
non-thermal---shock waves}

\maketitle

\begin{abstract}
In order to investigate the dispersion of the spectral break energy in
gamma-ray bursts,
we simulate internal shocks, including effects of
shell-splitting after shell collisions
and the Thomson optical depth due to electron--positron pairs
produced by synchrotron photons.
We produce pseudo observational data
and estimate the break energy.
If the distribution of initial Lorentz factors of shells
has only one peak,
many pulses with a break energy much lower than
the typical observed value should be detected,
while the effect of the Thomson optical depth
reduces pulses with a break energy of $>1$ MeV.
If the Lorentz factor distribution has multiple peaks,
the distribution of the break energy can be consistent with observations.
\end{abstract}


\section{Introduction}

\indent

Gamma-ray bursts (GRBs) are short ($\lesssim 10$ s) bursts of
low-energy $\gamma$-rays.
One of the most important characteristics of the GRB spectra 
is the existence of a typical break energy scale. 
The observed spectra of GRBs are approximated by a broken power 
law and the photon number spectra are approximated as 
$\sim \varepsilon^{-2}$ above the break energy and 
$\sim \varepsilon^{-1}$ below that.
The standard scenario for producing GRBs
is the dissipation of the kinetic energy of a
relativistic flow by relativistic shocks (see, e.g., a review by Piran 1999).
The rapid time variabilities observed require that the GRB
itself must arise from internal shocks within the flow.
The radiation is
emitted by relativistic electrons in the shocked region. The observations
are well-described by synchrotron emission \citep{coh97,wij99}.
\citet{kob97} shows that the internal shock
can reproduce the temporal structure of GRBs.
A large fraction of the kinetic energy could be converted to radiation
by the internal shocks \citep{kob01}.

The internal shock model is roughly represented by colliding shells.
The collisions are considered to occur within a wide range of radius $r$.
The energy efficiency argument
requires a large dispersion of the initial Lorentz factor $\Gamma$
\citep{kob01,bel00}.
In spite of the large dispersions of $r$ and $\Gamma$ in the model,
the apparent clustering of the break energy of the GRB spectra in the 50 keV--1 MeV
range is reported in the BATSE observation \citep{pre00}.
It is strange that pulses in a burst tend to have similar break energies.
The break energy distribution is in agreement with a log-normal distribution.
This distribution seems to suggest small dispersions of $r$ or $\Gamma$.

In this paper, based on numerical simulations,
we examine whether the standard internal shock model
can reproduce the narrow distribution of the break energy.
\citet{gue01} showed that the internal shock model
can reproduce the typical break energy.
However, the dispersion of the break energy has not been
explained so far.
In our simulation, as \citet{gue01} did, we include the effect of
the Thomson optical depth due to $e^\pm$ pairs produced
by synchrotron photons.
We treat the pair optical depth more precisely.
In addition, our improvements in the numerical simulations over those in \citet{gue01}
take into account shell splitting and the spectral energy band in the observation.
From our simulation we obtain some restrictions on GRB models.
In section 2 we explain our simulation method.
In section 3 the numerical results are given.
In section 4 we summarize our results.

\section{Method of Simulation}

In order to examine the distribution of the break energy,
we consider a spherical wind consisting of $N$ shells.
Each shell is characterized by the Lorentz factor $\Gamma_i$,
mass $M_i$, width $W_i$, and radius from the center $r_i$.
Following the evolution of the shells,
there will be numerous collisions between different shells.
For each collision we calculate the break energy,
optical depth, and flux.
In order to simplify our model we idealize the situation.
Some approximations are adopted in our formulation,
as discussed below.

\subsection{Break Energy}

\indent

In this subsection we explain our method to obtain the break energy.
Let us consider shells with different Lorentz factors,
$x\equiv\Gamma_{\rm r}/\Gamma_{\rm s} >1$. The relative Lorentz factor is $\Gamma_{\rm rel}
\sim (x+1/x)/2$. The rapid and slower shells are denoted by
subscripts r and s, respectively.
We assume that the widths of the shells are comparable in the
interstellar medium (ISM) rest frame as $W_i=W$.
Under these assumptions,
we can obtain the ratio of the baryon number density, $n_{\rm r}/n_{\rm s}$.
For equal mass shells ($M_i=$ const.) $n_{\rm r}/n_{\rm s}=1/x$,
while $n_{\rm r}/n_{\rm s}=1/x^2$
for equal energy shells ($M_i \Gamma_i=$ const.).
In this paper, $n$ and internal energy density $e$ are
measured in the fluids' rest frame.

When the rapid shell catches up with the slower one at
some radius $r$, the forward and reverse shocks form. The shock
conditions and the equality of pressures along the contact discontinuity 
\citep{sar95} yield
\begin{eqnarray}
\frac{(\Gamma_{\rm F}-1)(4 \Gamma_{\rm F}+3)}
{(\Gamma_{\rm R}-1)(4 \Gamma_{\rm R}+3)}=\frac{n_{\rm r}}{n_{\rm s}},
\label{jump}
\end{eqnarray}
where $\Gamma_{\rm F}$ and $\Gamma_{\rm R}$ are the Lorentz factors of 
the relative motion between the regions separated by the forward
shock and by the reverse shock, respectively.
The equality of the velocities along the contact discontinuity gives
$\Gamma_{\rm R}=\Gamma_{\rm F} \Gamma_{\rm rel}-\sqrt{(\Gamma_{\rm F}^2-1)
(\Gamma_{\rm rel}^2-1)}$.
In the case of equal mass shells or equal energy shells,
the solution for $\Gamma_{\rm R}$ depends on only the ratio of the Lorentz factors $x$.
The internal energy density in the shocked regions is given by 
\begin{eqnarray}
e=\frac{(\Gamma_{\rm R}-1) 
(4 \Gamma_{\rm R}+3) M_{\rm r} c^2}{4 \pi r^2  \Gamma_{\rm r} W},
\end{eqnarray}
where $M_{\rm r}$ is the mass of the rapid shell.

We now consider the synchrotron emission from shocked shells.
Since a rapid shell has a larger energy and a lower number density
in the cases of equal mass and equal energy,
the average energy per one electron in the reverse shock is larger
than that in the forward shock.
The reverse shock emission is more luminous and harder
than the forward shock emission. Therefore, we consider only the emission from 
the reverse shock. The shock is assumed to accelerate electrons in the shell
material into a power-law distribution, $N(\gamma_{\rm e}) \propto\gamma_{\rm e}^{-p} 
(\gamma_{\rm e} \ge \gamma_{\rm e,min})$. Assuming that constant fractions,
$\epsilon_B$  and $\epsilon_{\rm e}$, of the internal energy go into
the magnetic field and the electrons, respectively, one finds that 
the magnetic field and the typical random Lorentz factor of
electrons are given by $B^2\propto\epsilon_B e$ and $\gamma_{\rm e,min}
\propto \epsilon_{\rm e}(\Gamma_{\rm R}-1)$, respectively.

The synchrotron process is a very efficient radiation process. With the
strong magnetic field required to produce the observed gamma-ray, the
synchrotron cooling time is very short compared to the dynamical time of
the shock. In this fast cooling case \citep{sar98}, 
the photon number density distribution at the fluid rest frame is given by
\begin{equation}
\frac{d n_\gamma (\varepsilon)}{d \varepsilon}= \frac{p-2}{2 p-2} 
\frac{\epsilon_{\rm e} e}{\varepsilon_{\rm p}^2} \cdot
\left\{
\begin{array}{ll}
\left( \varepsilon/\varepsilon_{\rm p} \right)^{-3/2} 
& \quad \mbox{for $\varepsilon < \varepsilon_{\rm p}$}, \\
\left( \varepsilon/\varepsilon_{\rm p} \right)^{-(p+2)/2} 
& \quad \mbox{for $\varepsilon \geq \varepsilon_{\rm p}$}, 
\end{array}\right.
\label{spc}
\end{equation}
where the break energy $\varepsilon_{\rm p}$ is the typical energy of 
synchrotron photons emitted by electrons of
$\gamma_{\rm e,min}$. Since the shocked fluid is moving with the Lorentz
factor, $ \Gamma_{\rm m} \sim  \Gamma_{\rm r} (\Gamma_{\rm R}-
\sqrt{\Gamma_{\rm R}^2-1})$, the observed break photon  
energy is 
\begin{eqnarray}
\varepsilon_{\rm p}^{\rm obs} = 610 ~\epsilon_{\rm e}^2 F(x)
\Gamma_{\rm s, 2} \sqrt{\Sigma/\Delta} \quad \mbox{keV},
\label{ep}
\end{eqnarray}
where 
$F(x)=x^{1/2} (\Gamma_{\rm R}-1)^{5/2} (4 \Gamma_{\rm R}+3)^{1/2} 
(\Gamma_{\rm R}-\sqrt{\Gamma_{\rm R}^2-1})$. It behaves as $\sim x^2/4$ at $x\gg1$.
We have scaled the parameters as
$\epsilon_{B,-1}=\epsilon_B/0.1$, 
$M_{48}=M_{\rm r} c^2/10^{48}$ ergs, $r_{13}=r/10^{13}$ cm, 
$W_7=W/10^7$ cm, and $\Gamma_{\rm s, 2}=\Gamma_{\rm s}/100$. 
We adopt $p=2.5$ here and hereafter. Since some parameters
appear in the following formulae in the same combinations, we have 
defined two variables, a surface density 
$\Sigma = M_{48}/r_{13}^2$
and an effective comoving shell width 
$\Delta = W_7 \Gamma_{\rm s, 2}/\epsilon_{B, -1}$
for convenience.
In our simulation we neglect the effect of the cosmological redshift
on the break energy.

\subsection{Optical Depth Due to $e^\pm$ Pairs}

\indent

Shells are initially optically thick due to Thomson
scattering by electrons associated with baryons in the shell.
For a large optical depth $\tau$,
the radiative cooling time can be estimated as $\sim \tau l/c$,
where $l = W \Gamma_{\rm r}/(4 \Gamma_{\rm R}+3)$ is the shell width
in the comoving frame.
Comparing this with the cooling time due to the shell spreading $\sim
l/c$, the radiated energy is negligible for $\tau\gg1$.
Then, all of the internal energy is transformed back to kinetic energy \citep{kob01}.
Thus, we could not observe the emission from it, until it comes from outside of
the photosphere, where the optical depth $\tau_{\rm T}=2 \sigma_{\rm T}M/4 \pi r^2
m_{\rm p} \sim 0.7 \Sigma$ becomes unity.
This is a well-know result, but 
the estimate may undergo a significant change in the internal shock
model if we take into account $e^\pm$ pairs produced by the synchrotron
photons \citep{gue01}. In the fluid rest frame, the
break photon energy $\varepsilon_{\rm p}$ is much smaller than
the electron mass. However, the photon distribution extends
to high-energy as a power law. The pairs caused by the high energy
photons may contribute significantly to the optical depth.

When a photon of energy $\varepsilon$ interacts with a photon of $\varepsilon'$
at an incident angle $\theta$,
the cross section of the pair creation is written as
$\sigma_\pm=\sigma_{\rm T} f(y)$
\citep{ber82},
where
\begin{eqnarray}
f(y)=\frac{3}{16} (1-y^2) \left[
(3-y^4) \ln{\frac{1+y}{1-y}}-2 y (2-y^2) \right].
\end{eqnarray}
The dimensionless value $y$ is defined by $
y^2=1-(2 m_{\rm e}^2 c^4)/[\varepsilon \varepsilon' (1-\cos{\theta})]$.
The optical depth to pair-creation is given by
\begin{eqnarray}
\tau_{\gamma \gamma}(\varepsilon)&=&\int (1-\cos{\theta})
d\Omega \int_{\frac{2 m_{\rm e}^2 c^4}
{\varepsilon (1-\cos{\theta})}}^{\infty} d\varepsilon'
\frac{d n_\gamma (\varepsilon')}{d \varepsilon' d \Omega}
\sigma_{\gamma \gamma}(\varepsilon',\varepsilon) l \\
&=&
\sigma_{\rm T}l
\frac{4(p-2)}{(p-1)(p+4)} 
\frac{\epsilon_{\rm e}  e}{ \varepsilon_{\rm p}}
\left( \frac{\varepsilon \varepsilon_{\rm p}}
{m_{\rm e}^2 c^4} \right)^{p/2} C(p),
\label{tau0}
\end{eqnarray}
where
\begin{eqnarray}
C(p) \equiv \int_0^1 dy (1-y^2)^{p/2-1} y f(y).
\end{eqnarray}
The value of $C(p)$ is not sensitive to $p$, and $C(2.5) \sim 0.075$.
Since we use the power-law distribution
$dn_\gamma/d\varepsilon \propto \varepsilon^{-(p+2)/2}$ in the integration,
the optical depth is overestimated for photons with high energy
$\varepsilon \gtrsim 2 m_{\rm e}^2 c^4/\varepsilon_{\rm p}$, which interact
with photons mainly in the low-energy portion $dn_\gamma/d\varepsilon
\propto \varepsilon^{-3/2}$. However, the exact value is not important
for the high-energy photons when we estimate the number of $e^\pm$ pairs 
below, because the corrected optical depth is also large enough to
annihilate all of them.  For $p=2.5$ we obtain 
\begin{equation}
\tau_{\gamma \gamma}(\varepsilon)=3.3 ~\epsilon_{\rm e}^{3/2} G(x)
\Sigma^{9/8} \Delta^{-1/8} \left(\varepsilon/m_{\rm e}c^2\right)^{5/4},
\label{tau}
\end{equation}
where 
$G(x)=(\Gamma_{\rm R}-1)^{13/8}(4 \Gamma_{\rm R}+3)^{1/8}/x^{1/8}$.
Although the optical depth to pair-creation for photons with
the typical energy of $\varepsilon_{\rm p} \ll m_{\rm e} c^2$ is very small, it does not 
necessarily imply that the number of $e^\pm$ pairs is negligible for
Thomson scattering. 
Considering that photons with energy $\varepsilon$ interact mainly with
those with energy $\sim 2m_{\rm e}^2c^4/\varepsilon$ to produce pairs, we can
estimate the optical depth due to the pairs as  
\begin{equation}
\tau_{\pm} \sim 2\sigma_{\rm T}l
\int_{\sqrt{2}m_{\rm e}c^2}^\infty d \varepsilon 
\frac{d n_\gamma (\varepsilon)}{d \varepsilon} 
(1-e^{-\tau_{\gamma \gamma}} ).
\label{num}
\end{equation}

If we consider only the effect by the electrons accompanied by
baryons, the expanding ejecta becomes optically thin when the 
surface density $\Sigma \propto r^{-2}$ decreases down to $\sim 1$. However,
the $e^\pm$ pairs produced by the synchrotron photons generally increase the
optical depth significantly. Then, the ejecta is required 
to expand beyond $\Sigma\sim1$ to radiate. In figure 1 we plot the value
of the surface density $\Sigma_{\rm M}$, at which the ejecta becomes
optically thin, as a function of $x$.
In order to plot figure 1, we have numerically solved equation (\ref{jump}) and
integrated equation (\ref{num}). 
If we describe the system as an inelastic collision between two masses,
and half of the internal energy is converted to pairs
as \citet{gue01} assumed, $\Sigma_{\rm M}$ is estimated 
to be $\sim 0.017 \epsilon_{\rm e}^{-1}\sqrt{x}/(1+x-2 \sqrt{x})$.
This approximation generally gives a smaller value.

Since the break energy $\varepsilon_{\rm p}^{\rm obs}$ is
proportional to $\Sigma^{1/2}$, a collision occurring with larger $\Sigma$
radiates harder photons.
Then, the hardest synchrotron photons are emitted just above the radius
where $\Sigma=\Sigma_{\rm M}$
for given values of the parameters $x$, $\Delta$, and
$\Gamma_{\rm s}$.
We define the maximum break energy $E_{\rm M}$
as the break energy emitted from
the radius where $\Sigma=\Sigma_{\rm M}$.
The value of $E_{\rm M}$ is obtained from equation (\ref{ep}) with
$\Sigma=\Sigma_{\rm M}$.
Figure 2 depicts $E_{\rm M}$ as a function of $x$.
We can see that the photon pair-creation
effect makes $E_{\rm M}$ significantly smaller (solid lines) compared
to the case without the effect (dotted lines).
According to figure 2, it is difficult to obtain
a break energy of more than 1 MeV for $x \le 20$.
The upper limit of the break energy is determined by the Thomson
optical depth due to $e^\pm$ pairs,
and it is close to the typical observed break energy.
This result agrees with \citet{gue01}.

\begin{figure}
\begin{center}
\FigureFile(70mm,70mm){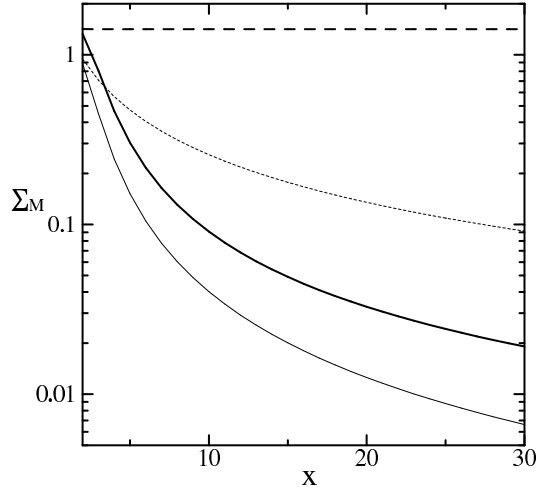}
\end{center}
\caption{Plot of $\Sigma_{\rm M}$ (solid line)
against $x$.
The thick and thin lines are
for equal-mass and equal-energy cases, respectively.
The dotted lines represent the values without pair-creation.
For equal mass the value is constant as $\Sigma_{\rm M}=1.4$.
Here, we assume $\Delta=1$ and $\epsilon_{\rm e}=0.5$.
}
\end{figure}

In the parameter region of our interest, the critical photon energy
$\varepsilon_\tau$, at which the optical depth $\tau_{\gamma
\gamma}=1$, is  larger than $\sqrt{2} m_{\rm e} c^2$. Then, we obtain the
approximation,
\begin{eqnarray}
\tau_{\pm} \sim 300 \epsilon_{\rm e}^{3} G^2(x) 
\Sigma^{9/4} \Delta^{-1/4},
\label{eq}
\end{eqnarray}
where we neglect a logarithmic factor of $\varepsilon_\tau$.
We adopt this approximation in our simulation.

\begin{figure}
\begin{center}
\FigureFile(70mm,70mm){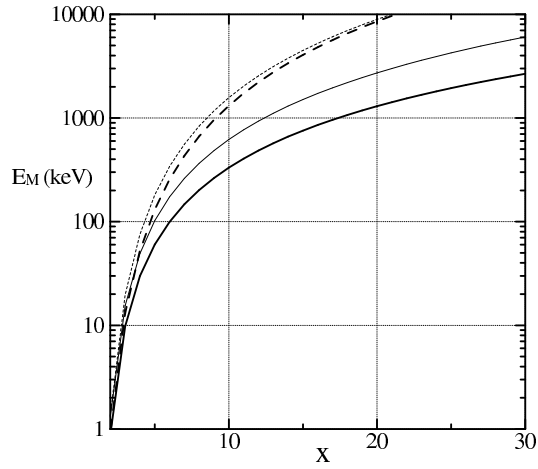}
\end{center}
\caption{Plot of $E_{\rm M}$ (solid line) against $x$.
The dotted lines are values obtained without the
process of pair-creation.
The thick and thin lines are
for equal-mass and equal-energy cases, respectively.
Here, we assume $\Delta=1$, $\epsilon_{\rm e}=0.5$, and $\Gamma_{\rm s, 2}=1$.
}
\end{figure}

\subsection{Sampling of Break Energy}

Setting the initial distributions of the Lorentz factor $\Gamma_i$
and separation between shells $L_{i}$,
we follow the evolution of the shells until there
are no more collisions, i.e., until the shells are
ordered by increasing value of the Lorentz factors.
For each collision, using equations (\ref{jump}), (\ref{ep}), 
and (\ref{eq}), we estimate the break energy and
the optical depth due to electron--positron pairs.
We approximate the radiation energy as
\begin{eqnarray}
E_{\rm rad}=\epsilon_{\rm e} \left[
M_{\rm r} (\Gamma_{\rm r}-\Gamma_{\rm m}')
+M_{\rm s} (\Gamma_{\rm s}-\Gamma_{\rm m}') \right],
\end{eqnarray}
where
\begin{eqnarray}
\Gamma_{\rm m}'=\sqrt{\frac{M_{\rm r} \Gamma_{\rm r}+M_{\rm s}
\Gamma_{\rm s}}{M_{\rm r}/\Gamma_{\rm r}
+M_{\rm s}/\Gamma_{\rm s}}},
\end{eqnarray}
as conventionally assumed \citep{kob01}.
After a fraction $\epsilon_{\rm e}$ of the internal energy
is emitted, the shells will split, transforming the remaining
internal energy back to kinetic energy.
In our simulation, the process of the shell splitting
is basically the same as in \citet{kob01}.
The introduction of shell splitting in the simulation increases
the number of collisions and the energy efficiency.
In this paper, for simplicity, we assume that each mass and width of the shell
are conserved before and after its collision.
If $\tau_{\pm} \geq 1$, we force $\epsilon_{\rm e}=0$.
Then the collisions of shells for $\tau_{\pm} \geq 1$ are similar to
the perfectly elastic collisions of pool balls.

In order to include the effect of the spectral energy band in observations,
we produce pseudo observational data from our simulation.
Using the spectrum of equation (\ref{spc}),
we estimate the flux between 20 and 2000 keV from the radiation energy
and break energy.
We neglect the effect of the spectral sensitivity
in BATSE instruments.
The effective area changes only by a factor of two
in most parts of the spectral energy band,
so that the effect of the sensitivity can be negligible.
From the estimated flux we write light curves by the same method as in \citet{kob97}.
The light curves are superpositions of many pulses emitted
from all collisions.
The light curve peaks are identified with the peak-finding
algorithm described by \citet{li96}.
A peak time $T_{\rm p}$ is identified,
if the peak photon flux $C_{\rm p}$, photon fluxes $C_1$ ($T_1<T_{\rm p}$),
and $C_2$ ($T_2>T_{\rm p}$)
satisfy $C_{\rm p}-C_{1,2}>N_{\rm var} \sqrt{C_{\rm p}}$
and there are no time bins between $T_1$ and $T_2$
with photon flux higher than $C_{\rm p}$.
We adopt $N_{\rm var}=0.3$.
The results do not strongly depend on $N_{\rm var}$ \citep{spa00}.
Each peak is identified as an ``observable pulse'' in our simulation.
The light curve valleys are identified as the minima between two
consecutive peaks.
We divide the light curves in time into some regions by valleys
and identify them as the observed duration time of each pulse.
One observable pulse may be the composition
of multiple pulses emitted from different collisions.
For each temporal region, by attaching a fluence weight,
we average the break energies of all pulses arrived during the period.
We adopt the average break energy as the ``observable break energy'' for
each temporal region.
This treatment is different from the method used in an actual observation.
In our simulation,
however, one pulse greatly overwhelms other dim pulses in most cases.
Therefore, our method is harmless when determining the break energy.

\section{Results of Simulation}

\subsection{Continuous $\Gamma$-Distribution}

In our simulation, the initial shells are assumed to have equal mass $M$ and width $W$.
The distributions of the initial Lorentz factor $\Gamma_i$ of each shell
and initial separation $L_{i}$ between two consecutive shells
are determined by the character of the central engine
on which we have little information.
In our simulation we assume that $\Gamma_i$
is to be distributed uniformly in logarithmic space between $\log{\Gamma_{\rm m}}$ 
and $\log{\Gamma_{\rm M}}$ (the number distribution
of shells is proportional to $1/\Gamma$).
The initial separation $L_{i}$
is also assumed to distribute in the same way
between $\log{W}$ and $\log{L_{\rm M}}$.
We perform 100 simulations with $N=50$, $\Gamma_{\rm m}=30$, $\Gamma_{\rm M}=3000$,
$W/c=10^{-2}$ s, $L_{\rm M}/c=1$ s,
$\epsilon_{\rm e}=0.6$, $\epsilon_B=0.1$,
and a total initial kinetic energy of $E_{\rm iso}=10^{52}$ erg.
First we present the results, neglecting the effect of $\tau_{\pm}$.
Figure 3 shows a histogram of the break energy
for all collisions, neglecting the spectral band in the observation.
The break energies widely distribute from the optical range
to 10 MeV.
This result apparently contradicts the observation.
We then take into account the Thomson optical depth.
As is shown in figure 4, the break energies above $\sim 1$ MeV
are suppressed, as we have speculated in subsection 2.2.
However, there remain many pulses with a low break energy.
The effect of the Thomson optical depth reduces
only high-break pulses.

\begin{figure}
\begin{center}
\FigureFile(70mm,70mm){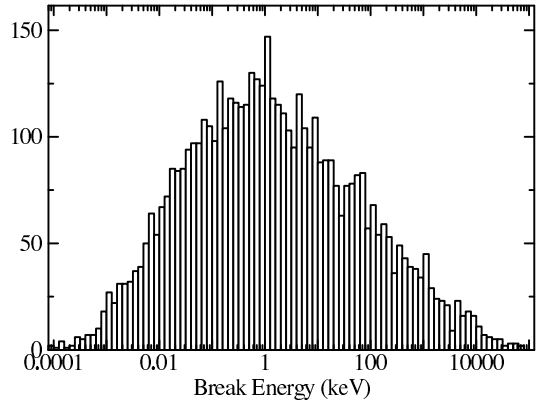}
\end{center}
\caption{Distribution of the break energy for all collisions.
The effect of the Thomson optical depth is neglected.
}
\end{figure}

However, we can not observe all of the pulses emitted from all collisions.
A dimmer pulse may be overwhelmed by other brighter pulses.
Since a dimmer pulse is expected to have a lower break energy,
actual observations reduce the number of pulses with a low break energy.
In addition, a large energy fraction of
emission with a very low break energy may be out of the spectral
energy band in the observation.
Therefore, we mimic the BATSE observation by producing light curves
from the simulations, as explained in subsection 2.3,
and define the observable pulses and their break energies.
In figure 4, we show the result obtained from the pseudo observational data.
Samples with lower break energy diminish drastically.
In figure 5 we magnify the result around the BATSE band.
Although the peak of the distribution is around the lowest limit of the BATSE band,
the dispersion is significantly larger than the observation.
There are many pulses with a break energy of $< 20$ keV,
which should be observed as ``no break'' pulses.
In the spectral energy band only high-energy tails
of bright spectra of such pulses are detected, and they make significant
peaks in the light curve.

\begin{figure}
\begin{center}
\FigureFile(70mm,70mm){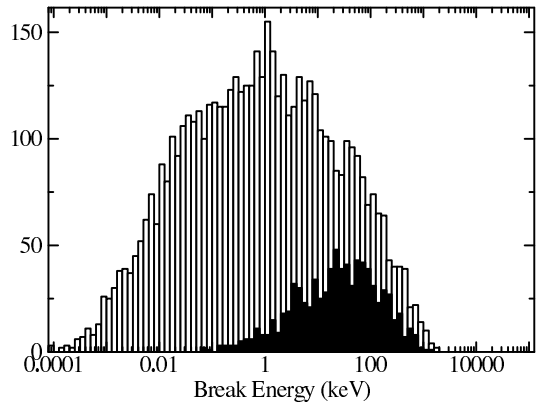}
\end{center}
\caption{Distribution of the break energy for all collisions (white)
and observable pulses (black).
The effect of the Thomson optical depth is taken into account.
}
\end{figure}

\begin{figure}
\begin{center}
\FigureFile(40mm,40mm){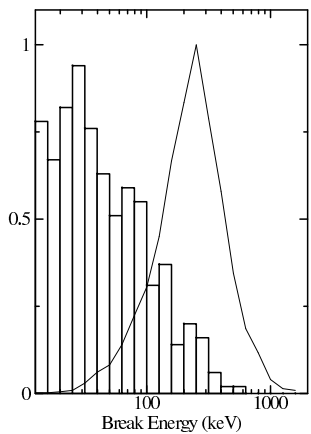}
\end{center}
\caption{Same as figure 4 for observable pulses.
Here we include the cosmological redshift effect, assuming $z=1.5$.
The solid line is the BATSE observation.
There are many other pulses outside of this energy band.
They should be counted as pulses whose spectra have no break energy.
}
\end{figure}

Regarding the observable pulses
in figures 6 and 7, we show the relations between the break energy
and the collision radius $r$,
and between the break energy and the ratio of the Lorentz factors $x$, respectively.
The collision radius and ratio $x$ are obtained averaging
in the same way as the break energy.
According to these figures,
the break energy is mainly determined by the ratio $x$ rather than $r$.
Collisions with a smaller $x$ produce lower break energies,
which cause a large dispersion of the break energy.
The lower limit of the collision radius in figure 6 is determined
by the Thomson optical depth.

\begin{figure}
\begin{center}
\FigureFile(100mm,70mm){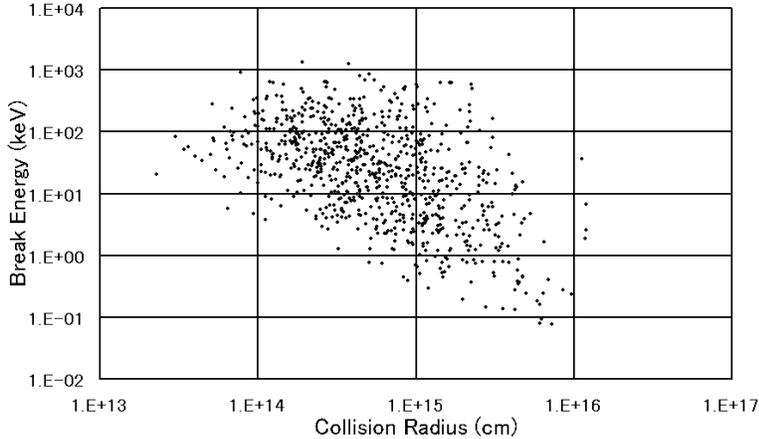}
\end{center}
\caption{Relation between the break energy and collision radius.
}
\end{figure}

\begin{figure}
\begin{center}
\FigureFile(100mm,70mm){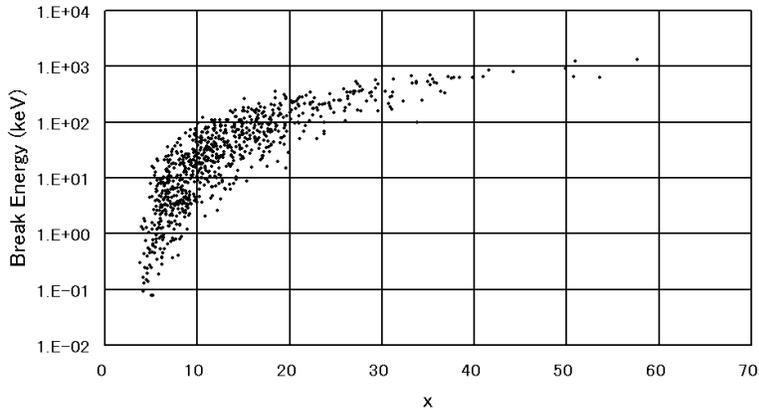}
\end{center}
\caption{Relation between the break energy and the ratio $x$.
}
\end{figure}

In this model, the fraction of the emission energy to the total
kinetic energy is $16.9 \pm 2.0$\%.
As \citet{kob01} claimed, the introduction of shell splitting
increases the energy efficiency compared with that of \citet{gue01}
(typically a few percent).
In the pseudo BATSE band, the energy efficiency is $2.9 \pm 0.9$\%.
About four-times radiation energy in the BATSE band is emitted
outside of the BATSE band.

\subsection{Bimodal $\Gamma$-Distribution}

We have simulated for other types of distributions of $\Gamma$.
However, as long as the $\Gamma$-distribution is continuous
and has only one peak,
the results do not change basically.
The models predict many ``no break'' pulses.
Let us consider a $\Gamma$-distribution that
has a maximum at $\Gamma=\Gamma_{\rm p}$.
If we randomly choose two shells in this distribution,
the most probable $\Gamma$s of two shells are around $\Gamma_{\rm p}$.
Therefore, we cannot avoid numerous collisions with $x \sim 1$,
which leads to low break energies.
Of course, a too small $x$ leads to too dim emission.
However, emission from a marginal value of $x$ ($\sim 5$)
is luminous enough in spite of the small break energy.

If the $\Gamma$-distribution has two peaks,
the probability function of $x$ may have a maximum at $x \gg 1$.
As one example, we make a simulation for a bimodal $\Gamma$-distribution:
one half of the shells have $\Gamma=30$ and the other half have $\Gamma=3000$ initially.
The parameters are common to those described in the former subsection.
The result is shown in figure 8.
Since collisions with $\Gamma=30$ and $\Gamma=3000$ are dominant events in this case,
break energies distribute narrowly.
The dispersion of the break energy is consistent with the observation.
It is trivial that we can adjust the peak of the distribution
to the observation more closely by changing the parameter $\epsilon_{\rm e}$.

The fraction of the emission energy to the total
kinetic energy is $65.1 \pm 6.6$\%.
In the pseudo BATSE band, the energy efficiency is $18.7 \pm 2.5$\%.
The bimodal distribution is advantageous to the energy efficiency too.

\begin{figure}
\begin{center}
\FigureFile(40mm,40mm){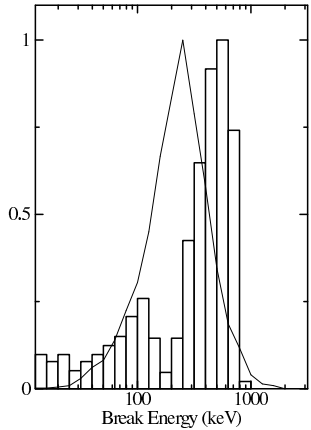}
\end{center}
\caption{Distribution of the observable break energy for the bimodal distribution.
The effect of the Thomson optical depth is taken into account.
Here, we include the cosmological redshift effect assuming $z=1.5$.
The solid line is the BATSE observation.
}
\end{figure}

\section{Conclusions and Discussion}

Following the standard scenario of GRB,
we simulate internal shocks, including the effects of
shell splitting and the Thomson optical depth due to electron--positron pairs
produced by synchrotron photons.
We produce pseudo observational data
and estimate the break energy.
The effect of the Thomson optical depth
reduces the pulses with a break energy of $>1$ MeV, which is consistent with \citet{gue01}.
The shell-splitting effect increases the energy efficiency.
However, many ``no break'' pulses should be observed in the case of a one-peak,
continuous $\Gamma$-distribution.
Even if we alter some assumptions (shell splitting, equal shell mass, etc.)
in our model, the qualitative result is basically the same.
Within our simple method, ``no break'' pulses are unavoidable,
though we take into account the spectral energy band in the observation.
Collisions with a small ratio $x$ cause ``no break'' pulses.
\citet{pre00} insists that the small dispersion of the break energy
is not due to the observational selection effects.
However, we should henceforth consider the possibility of selection effects
on the BATSE observation.

The $\Gamma$-distribution is determined by the character of the central engine.
If the initial $\Gamma$-distribution is discrete,
or has multiple peaks, like the bimodal distribution,
the distribution of the break energy can be consistent with the observation.
The bimodal distribution is favorable for the energy efficiency.
However, we do not know if such a distribution is realistic or not.

Judging from figure 7, if we choose pulses emitted from collisions with $x \gtrsim 10$,
most of the observable break energies are in the BATSE band.
Therefore, if some microscopic processes prohibit collisions with $x \lesssim 10$
from emitting photons,
the small dispersion of the break energy can be reproduced.
One of the candidates for such processes is in the electron acceleration mechanism,
which is not well understood.
In the standard GRB model, a large 
fraction of the kinetic energy carried by protons is efficiently 
converted into that of relativistic electrons in the shocked region.
However, this premise has not been proven theoretically. 
It is apparent that the Coulomb interaction cannot transport the internal 
energy of heated protons 
into electrons to achieve energy equipartition, because
the time scale of the Coulomb interaction is much longer
than the dynamical time scale.
In our model, $x=10$ corresponds to $\Gamma_{\rm R}=2.57$.
Therefore, the threshold $x \simeq 10$ is a border between mild-relativistic and
ultra-relativistic shock.
If the energy of protons
can be efficiently transported into electrons only for ultra-relativistic shock,
the distribution of the break energy can be explained.
\citet{hos02} proposed a new acceleration mechanism: shock surfing acceleration.
This mechanism effectively accelerates only electrons in the cases of a strong shock.
Such a study might explain the break energy in the future.

\vspace{1.5cm}

We would like to thank T. Sakamoto for providing information 
about the BATSE instruments. We also thank R. Preece for
useful discussions, as well as providing and helping with the 
some of the BATSE data. K.A. is supported by the Japan Society
for the Promotion of Science. S.K. thanks support through the 
Center for Gravitational Wave Physics, which is funded by NSF 
under cooperative agreement PHY 01-14375.

\end{document}